\documentclass[aps,twocolumn,pra,showpacs,floatfix]{revtex4}
\usepackage{epsfig}
\usepackage{graphicx}
\usepackage{dcolumn}
\usepackage{amssymb,amsmath}
\usepackage{mathrsfs}
\begin{document}

\title{The Os$^{16+}$ and Ir$^{17+}$ ions as candidates for accurate optical clock sensitive to physics beyond standard model.}

\author{V. A. Dzuba, V. V. Flambaum}

\affiliation{School of Physics, University of New South Wales, Sydney 2052, Australia}

\begin{abstract}
We perform detailed calculations of the electronic structure of the Os$^{16+}$ ion and demonstrate that it has several metastable states which can be used for very accurate optical clocks. The clocks are highly sensitive to manifestations of the physics beyond standard model, such as time variation of the fine structure constant $\alpha$, interaction with scalar and pseudoscalar (axion) dark matter  fields, local Lorentz invariance and local position invariance violations,
 and interaction of atomic electrons with nucleus mediated by new boson. The latter can be studied by analysing King plot for isotope shifts  and its possible non-linearities since Os has 5 stable isotopes with zero nuclear spin.
 Similar calculations for the Ir$^{17+}$ ion spectra %and some minor amendments to the interpretation of the experimental data 
 demonstrate good agreement between theory and experiment. This helps to validate the method  of the calculations and demonstrate that both ions are excellent candidates for the search of new physics.
 %share many common features in terms of searching for new physics. 
\end{abstract}

%\pacs{31.15.A-,11.30.Er}

\maketitle

\section{Introduction}

It was suggested in Refs.~\cite{HCI} to use highly charged ions (HCI) to search for optical transitions highly sensitive to the time variation of the fine structure constant $\alpha$. 
The idea is based on the fact of the {\em level crossing} \cite{crossing}. Usually intervals between electron energy levels are very large in HCI compare to neutral atoms.  However, due to different level ordering in neutral atoms and hydrogen-like ions, the energy interval between states of different configurations, drawn as a function of the ionisation degree $Z_i$, must cross at some point, brining the energy interval into the optical region.
Since states of different configurations have different dependence on the value of the fine structure constant $\alpha$, the energy intervals are very sensitive to the variation of $\alpha$. The sensitivity is proportional to $ Z^2 ( Z_i +1)^2$ and strongly depends on the electron orbital angular momentum. The largest sensitivity can be found in electron transitions in heavy ions which in singe-electron approximation can be described as $s_{1/2}$ - $f_{5/2},f_{7/2}$ or $p_{1/2}$ - $f_{5/2},f_{7/2}$ ($s$-$f$ or $p$-$f$) transitions~\cite{HCI,crossing,Bekker}.
Use of metastable states brings additional advantage of potentially very high accuracy of the measurements  typical for atomic optical clocks.
The accuracy for HCI clocks can be even higher than that for optical clocks in neural atoms due to the fact that states of HCI are less sensitive to perturbations due to compact size of HCI, small polarisability
and large energies of excitations \cite{hci3}.   

\begin{figure}[tb]
	\epsfig{figure=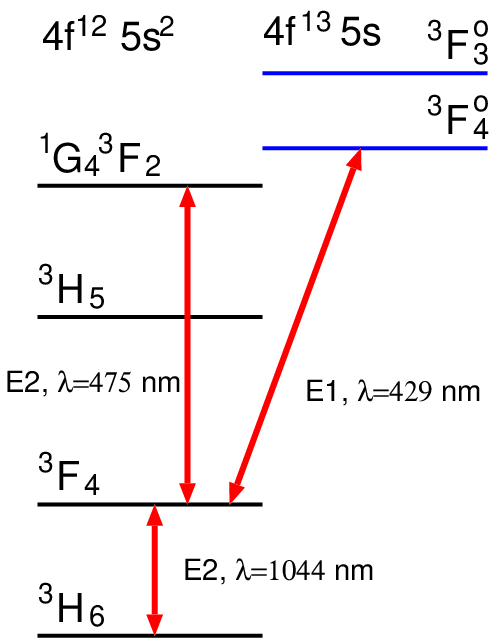,scale=0.7}
	\caption{Energy level diagram for the Os$^{16+}$ ion. Possible clock transitions within  even configuration and $s-f$ transition  are shown as red double arrows.}
	\label{f:Os}
\end{figure}

A number of candidate systems were suggested in earlier works~\cite{hci1,hci2,hci3,hci4} (see also reviews~\cite{rev1,rev2} and references therein).
Experimental studies were performed for the Ho$^{14+}$~\cite{Ho14-1,Ho14-2} and Ir$^{17+}$~\cite{Ir17} ions. Further work is in progress~\cite{rev1,rev2}.
In present work we study the Os$^{16+}$ ion. It has some important features which make it attractive candidate for experimental study.
It has several metastable states which can be used for clock transitions. At least one transition is  $s-f$ transition, so that it is very sensitive to the variation of the fine structure constant  $\alpha$ and dark matter field which may be a source of such variation \cite{Arvanitaki,Stadnik,Stadnik2}. Other transitions are less sensitive to $\alpha$ variation and can serve as {\em anchor} lines. In addition, they are sensitive to other manifestations of new physics such as local Lorentz invariance and local position invariance violations, etc.
The energy diagram for Os$^{16+}$ ion is presented on Fig.~\ref{f:Os}. This diagram is the result of the calculations in the present work.
%Experimental energies of the Os$^{16+}$ ion are not known.
Experimental energy intervals between states of different configurations are not known,

\begin{table}
\caption{\label{t:OsIs}
A list of stable isotopes of Os with zero nuclear spin. Parameters $\beta$ of the  quadrupole deformation of proton distribution are taken from Ref.~\cite{ddpc}.}
\begin{ruledtabular}
\begin{tabular}   {cccccc}
%\hline
$A$      & 184 & 186 & 188 & 190 & 192 \\
$\beta$ & 0.281 & 0.257 & 0.223 & 0.185 & 0.164 \\
\end{tabular}			
\end{ruledtabular}
\end{table}

The Os$^{16+}$ ion is similar to the Ir$^{17+}$ ion studied before~\cite{Ir17,Ir17t}. However, it has important advantage of having five stable isotopes with zero nuclear spin (Ir has none). It makes this ion suitable for searching for new interactions via looking at possible non-linearities of King plot~\cite{KP1,KP2}.
The minimum requirements for such study include having two clock transitions and four stable isotopes. Isotopes with zero nuclear spin have further advantage of having no hyperfine structure which complicates the analysis of the isotope shift. Table~\ref{t:OsIs} lists five stable isotopes of Os which has zero nuclear spin. It also presents the parameters $\beta$ of nuclear quadrupole deformation. These parameters come from nuclear calculations~\cite{ddpc}. Nuclear deformation can lead to the non-linearities of King plot~\cite{deform,Yb+} presenting important systematic effect in search for new interactions. Note however that the parameters of deformation have similar values for all stable isotopes (see Table~\ref{t:OsIs}).
This means that significant cancellation of the effect of deformation is possible in the isotope shift.

Finally, the Os$^{16+}$ and Ir$^{17+}$ ions are suitable for searching the effects of local Lorentz invariance (LLI) and local position invariance (LPI) violation, since these effects are strongly enhanced in HCI~\cite{Shaniv}. It was argued in Refs.~\cite{Shaniv,Nature}  that to have enhanced value of the LLI violation one needs to have long-living state (e.g, ground) with large value of the total electron momentum $J$ ($J \geq 2$) and large values of the matrix element for the LLI violating operator.
Such states can often be found in open $4f$-shells.
All these conditions are satisfied for metastable states of the Os$^{16+}$ and Ir$^{17+}$ ions, including ground state. 
%This feature is not unique to the Os$^{16+}$ ion.
%The same is true for all HCI with open $4f$ (or $5f$) shell.
%The values of the matrix elements of the LLI violating operator for metastable states of  the Os$^{16+}$ ion are larger than those in the Yb$^+$ ion.
%Note that an excited metastable state of Yb$^+$ ion was considered in Ref.~\cite{Nature}.
The  situation with the LPI violating effect in HCI is similar to those of the LLI violating effect.

\section{Method of calculations}

The Os$^{16+}$ ion has an open $4f$ shell. Its ground state configuration is [Pd]$4f^{12}5s^2$, while we study also the states of the [Pd]$4f^{13}5s$ configuration. The ion has fourteen electrons in open shells. This number is too large for standard configuration interaction (CI) calculations.
In present work we use the configuration interaction with perturbation theory (CIPT) method~\cite{cipt} especially developed for such systems.
The method was used for many atoms and ions with open $d$ or $f$ shells (see, e.g.~\cite{cipt1,cipt2,cipt3}) and it is proved to be very useful.
The best result are achieved for systems with almost empty or almost full open shells. On the other hand, the systems with half filled $d$ or $f$ shell are the most difficult for calculations. Since the accuracy for different systems is different, it is always useful  to do some additional tests for similar systems where experimental data or other calculations are available. 
In this work we study the Os$^{16+}$ and Ir$^{17+}$ ions. Both ions have similar electronic structure (the configurations are the same but states go in different order). There are experimental data on a number of transition energies for both ions~\cite{Ir17}. The same paper also presents several different calculations. However, the data on Ir$^{17+}$ ion is more complete. Experimental data includes energies for some E1 transitions between states of different configurations, while the data for Os$^{16+}$ contains only energies of M1 transitions between states of the same configuration.
The E1 transitions present special interest because of their sensitivity to the variation of the fine structure constant. The initial idea of using Ir$^{17+}$ ion comes from the finding that the energy of the E1 transition is in optical region~\cite{hci1}.
%There is no experimental data on the energy levels of the Os$^{16+}$ ion. However, many transition frequencies has been measured in the Ir$^{17+}$ ion which has similar electronic structure (the configurations are the same but states go in different order). 
Very advanced calculations by a different method~\cite{Ir17t} are also available for Ir$^{17+}$. Therefore, we use calculations for Ir$^{17+}$ to check the accuracy and applicability of the CIPT method to the Os$^{16+}$ ion.

\subsection{Calculation of energies}

The wave function for fourteen external electrons of the Os$^{16+}$ ion in the CIPT method is presented as an expansion over single-determinant many-electron basis functions
\begin{eqnarray}\label{e:psi}
\Psi(r_1, \dots, r_{N_e}) &=& \sum_{i=1}^{N_1} c_i \Phi_i(r_1, \dots, r_{N_e}) \\
&+& \sum_{i=N_1+1}^{N_2} c_i \Phi_i(r_1, \dots, r_{N_e}). \nonumber
\end{eqnarray}
Here $N_e$ is the number of external electrons ($N_e = 14$ in our case).
The expansion is divided into two parts. It is assumed that first few low-energy (the energy is related to the basis state by $E_i = \langle \Phi_i| \hat H^{\rm CI} | \Phi_i \rangle$) terms present good approximation for the wave function, while huge number ($N_2 \gg N_1$) of remaining high-energy terms is just a small correction. Calculations start from the relativistic Hartree-Fock (RHF) method applied to the open-shell ion. To make sure that first part of expansion (\ref{e:psi}) presents good approximation for the wave function, the electron configuration in the RHF calculations should coincide with one of the configurations of interest. In our case these are the $4f^{12}5s^2$ or $4f^{13}5s$ configurations. Changing initial choice from one configuration to another changes energy intervals between states of different configurations by few thousand cm$^{-1}$ without changing the order of the states. 
We have chosen the $4f^{13}5s$ configuration in the RHF because this gives good results for Ir$^{17+}$ (see next section).
The RHF Hamiltonian has the form
\begin{equation}\label{e:RHF}
\hat H^{\rm RHF} = c \mathbf{\alpha}\cdot \mathbf{p} + (\beta-1)mc^2 + V_{\rm nuc} + V_{\rm Breit} + V_{\rm QED} + V_e,
\end{equation}
where $c$ is speed of light, $\mathbf{\alpha}$ and $\beta$ are Dirac matrixes, $\mathbf{p}$ is electron momentum, $V_{\rm nuc}$ is nuclear potential obtained by integrating Fermi distribution of nuclear charge, $V_{\rm Breit}$ is the operator of Breit interaction which includes magnetic interaction and retardation~\cite{Breit} in n zero-frequency approximation (see, e.g.~\cite{Breits}),
  $V_{\rm QED}$ is the potential which simulates quantum electrodynamic corrections~\cite{QED},
$V_e$ is electron self-consistent RHF potential with contributions from all 60 electrons of the Os$^{16+}$ ion including the electrons of the $4f^{13}5s$ configuration.
On next stage the single-electron basis is calculated in the field of frozen core using the B-spline technique~\cite{B-spline}.

Then, applying the standard CI technique and neglecting the off-diagonal matrix elements between high-energy states on gets the CIPT equation 
\begin{equation}\label{e:cipt}
%\left( \langle i |\hat H^{\rm CI} | j \rangle +\sum_{m}^{N_2} \frac{\langle i |\hat H^{\rm CI} | m \rangle\langle m |\hat H^{\rm CI} | j \rangle}{E-E_m} - E\delta_{ij}\right) \mathbf{X}= 0.
\left[ \langle i |\hat H^{\rm CI} | j \rangle +\sum_{m}^{N_2} \frac{\langle i |\hat H^{\rm CI} | m \rangle\langle m |\hat H^{\rm CI} | j \rangle}{E-E_m}\right]\mathbf{X} = E\mathbf{X}.
\end{equation}
Here $\mathbf{X}$ is the vector of unknown expansion coefficients, $\mathbf{X} = (c_1, \dots , c_{N_1})$.
Indexes $i,j,m$ numerate many-electron basis states $|\Phi\rangle$, indexes $i$ and $j$ run from 1 to $N_1$, index $m$ runs from $N_1+1$ to $N_2$. 
Operator $\hat H^{\rm CI}$ is the CI Hamiltonian 
\begin{equation}\label{e:hci}
\hat H^{\rm CI} = \sum_{i=1}^{N_e} \hat H_{1i} + \sum_{i>j}^{N_e} \frac{e^2}{|\mathbf{r}_i - \mathbf{r}_j|},
\end{equation}
where $\hat H_{1i}$ is the single electron part of the Hamiltonian similar to (\ref{e:RHF}) but with $V_e$ replaced by $V_{\rm core}$.
Only core electrons (up to the $4d$ shell) contribute the the $V_{\rm core}$ potential, while all ionic electrons (including the $4f^{13}5s$ configuration) contribute the the $V_e$ potential.
We do not include $V_{\rm Breit}$ into the CI Hamiltonian because Breit interaction between valence electrons is a small correction which is much smaller than the uncertainty coming from other sources. The inclusion of Breit and QED corrections is important on the RHF stage only since it involves all atomic electrons including inner-shell ultra-relativistic electrons.
%\begin{equation}\label{e:h1ci}
%\hat H^{\rm RHF} = c \mathbf{\alpha}\cdot \mathbf{p} + (\beta-1)mc^2 + V_{\rm nuc} + V_{\rm core}.
%\end{equation}
%Here only core electrons (up to $4d$) contribute the the $V_{\rm core}$ potential.

The typical values of the $N_1$ and $N_2$ parameters for different states of Os$^{16+}$ are presented in Table~\ref{t:CIPT}.
The value of $N_1$ is the size of the effective CI matrix. Note, that it is always small. The main challenge of the method is the calculation of the second-order correction containing the huge number of terms ($N_1\times N_2$, see Table~\ref{t:CIPT} for the values).

The energy $E$ in (\ref{e:cipt}) is the energy of the state to be found from solving the CIPT equations. It presents in both, left and right-hand sides of the equation. This means that iterations are needed to solve the equations. Iterations can start from solving the CIPT equations (\ref{e:cipt}) without the second-order correction. In most cases less than ten iterations are sufficient for full convergence. 

\begin{table}
\caption{\label{t:CIPT}
Parameters of the CIPT calculations for the Os$^{16+}$ ion. $J^p$ stands for the total angular momentum and parity.
%Two non-relativistic configurations are used in each case, even and odd parity, to generate states in the effective CI matrix.
One even configuration ($4f^{12}5s^2$), and two odd configurations ($4f^{13}5s$ and $4f^{12}5s5p$) are used to generate states in the effective CI matrix.
Second odd configuration is added to allow electric dipole transitions between states of even and odd configurations (see Fig.~\ref{f:e1}).
$N_c$ is the corresponding number of relativistic configurations, $N_1$ is the corresponding number of states with given $J^p$; $N_1 \times N_1$ is the size of the effective CI matrix; $N_2$ is the number of terms in the second-order correction (second term in (\ref{e:cipt})).
}
\begin{ruledtabular}
\begin{tabular}   {rrrr}
\multicolumn{1}{c}{$J^p$}&
\multicolumn{1}{c}{$N_c$}&
\multicolumn{1}{c}{$N_1$}&
\multicolumn{1}{c}{$N_2$}\\
\hline
$3^-$ & 8 & 26 & $\sim 3 \times 10^5$ \\
$4^-$ & 7 & 24 & $\sim 3 \times 10^5$ \\
$2^+$ & 3 & 3 & $\sim 6 \times 10^6$ \\
$3^+$ & 3 & 1 & $\sim 1.5 \times 10^7$ \\
$4^+$ & 3 & 3 & $\sim 4 \times 10^6$ \\
$5^+$ & 3 & 1 & $\sim 1.0 \times 10^7$ \\
$6^+$ & 3 & 2 & $\sim 4 \times 10^6$ \\
\end{tabular}			
\end{ruledtabular}
\end{table}

\subsection{Calculation of matrix elements}

\label{s:RPA}

To calculate matrix elements of transitions between states and energy shifts due to different effects which were not included in the calculations of energy we use the time-dependent Hartree-Fock (TDHF) method~\cite{TDHF} which is equivalent to the well known random phase approximation (RPA).
The RPA equations have the form
\begin{equation}\label{e:RPA}
(\hat H^{\rm RHF} - \epsilon_c)\delta \psi_c = -(\hat F + \delta V_e^F)\psi_c.
\end{equation} 
Here $\hat H^{\rm RHF}$ is given by (\ref{e:RHF}), index $c$ numerate single-electron states of the ion (the same as in the RHF calculations), $\hat F$ is the operator of external field, $\delta \psi_c$ is the correction to the wave function due to external field, $ \delta V_e^F$ is the correction to the self-consistent RHF potential caused by the change of all ionic states. 
%The list of operators of external field considered in present paper is presented in Table~\ref{t:F}.

The RPA equations are solved self-consistently for all RHF states of the ion. Then transition amplitude is given by
\begin{equation}\label{e:Tij}
T_{ij} = \langle \Psi_i |\sum_{m=1}^{N_e}(\hat F + \delta V_e^F)_m| \Psi_j \rangle,
\end{equation} 
where $|\Psi_i \rangle$, $|\Psi_j \rangle$ come from solving the CIPT equations (\ref{e:cipt}). For energy shifts $i=j$ in (\ref{e:Tij}).

%\begin{table}
%\caption{\label{t:F} Operators of external field ($F$) used in this work.
%$J^p$ stands for the rank and parity of the operator; $C_{km}$ are the normalised spherical functions.}
%\begin{ruledtabular}
%\begin{tabular}   {llll}
%\multicolumn{1}{c}{Name}&
%\multicolumn{1}{c}{Notation}&
%\multicolumn{1}{c}{$J^p$}&
%\multicolumn{1}{c}{Expression}\\
%\hline
%\multicolumn{4}{c}{Transitions}\\
%Electric dipole & E1 & $1^-$ & $rC_{1m}(\theta,\phi)$ \\
%Magnetic dipole & M1 & $1^+$ & $C_{1m}(\theta,\phi)$ \\
%Electric quadrupole & E2 & $2^+$ & $r^2C_{2m}(\theta,\phi)$ \\
%\multicolumn{4}{c}{Energy shifts}\\
%Field shift & FS & $0^+$ & $\delta V_{\rm nuc}/\Delta \langle r^2 \rangle$ \\
%EEP violation & $\hat K$ & $0^+$ & $c\gamma_0\gamma^ip_i/2$ \\
%LLI violation\footnotemark[1]  & $T_0^{(2)}$ & $2^+$ & $c\gamma_0(\gamma^ip_i-3\gamma^3p_3)$ \\
%\end{tabular}
%\footnotetext[1]{Energy shifts between states with different values of $J_z$.} 			
%\end{ruledtabular}
%\end{table}

\begin{figure}[tb]
	\epsfig{figure=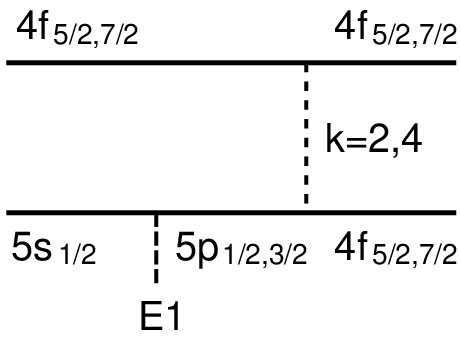,scale=0.8}
	\caption{Dominating contributions to the electric dipole (E1) transition between states of even ($4f^{12}5s^2$) and odd ($4f^{13}5s$) configurations.
	The transition is due to the mixing of the odd states of the $4f^{13}5s$ and $4f^{12}5s5p$ configurations.}
	\label{f:e1}
\end{figure}

Note that in some cases one needs to include more terms in the first part of the wave function expansion (\ref{e:psi}) to allow for non-zero value of transition amplitudes.
For example, the electric dipole (E1) transition cannot go directly between states of the $4f^{13}5s$ and $4f^{12}5s^2$ configurations because in this case it would correspond to the $s-f$ single-electron transition which is forbidden by the selection rules. One needs to add at least the $4f^{12}5s5p$ configuration to mix with the 
$4f^{13}5s$ configuration. Then the E1 transition amplitude is given by the diagram on Fig.~\ref{f:e1}.

\subsection{Energy levels of Ir$^{17+}$; discussion of accuracy}

The Ir$^{17+}$ ion was proposed for the measurements in \cite{hci1} and studied both experimentally and theoretically (see \cite{Ir17,Ir17t} and references therein). It is very similar to the Os$^{16+}$ ion studied in present work. It has the same number of external electrons forming the same configurations and many other similarities. It is natural to expect that the accuracy of the calculations for both ions is very similar too. Therefore, we perform exactly the same calculations for both ions and compare our results with experiment and previous calculations. Our first calculations for the Ir$^{17+}$ ion~\cite{hci1} used a different method which was less accurate, especially for the energy intervals between states of different configurations.
It had fitting parameters which would allow to fix the interval when experimental value is known. However, in absence of the experimental data the predictions were not very accurate. In contrast, our present method is fully {\em ab initio} and produces more accurate result. There were many other calculations, see~\cite{Ir17} and references therein. The most sophisticated and accurate calculations were reported in Ref.~\cite{Ir17t}. The agreement with experiment was very good.
Therefore, we compare our present results only with experiment and calculations of Ref.~\cite{Ir17t}.

\begin{table}
\caption{\label{t:IrE}
Energy levels and $g$-factors of the Ir$^{17+}$ ion. Comparison with previous calculations and experiment.
Experimental energies are obtained from measured transition energies~\cite{Ir17}, see also Tables~\ref{t:Irwm1} and \ref{t:Irwe1}.}
\begin{ruledtabular}
\begin{tabular}   {ll rrcrr}
\multicolumn{1}{c}{Config.}&
\multicolumn{1}{c}{Term}&
\multicolumn{4}{c}{Energies [cm$^{-1}$]}&
\multicolumn{1}{c}{$g$-factors}\\
&&\multicolumn{1}{c}{Ref.\cite{hci1}}&
\multicolumn{1}{c}{Ref.\cite{Ir17t}}&
\multicolumn{1}{c}{Expt.}&
\multicolumn{2}{c}{Present work}\\
\hline
\multicolumn{7}{c}{Odd states}\\
$4f^{13}5s$ & $^3$F$^{\rm o}_4$ &     0 &     0 &       &     0 & 1.2500 \\
            & $^3$F$^{\rm o}_3$ &  4838 &  4777 &       &  4647 & 1.0515 \\
            & $^3$F$^{\rm o}_2$ & 26272 & 25186 &       & 25198 & 0.6667 \\
            & $^1$F$^{\rm o}_3$ & 31492 & 30395 & 30359 & 30167 & 1.0318 \\
\multicolumn{7}{c}{Even states}\\
$4f^{14}$    & $^1$S$_0$  &  5055 &   12382 &  &    7424 &   0.0000  \\
&&&&&\\
%                            old     Safr   Expt.     new   
$4f^{12}5s^2$&$^3$H$_6$  & 35285 &  30283 &       &  29695 & 1.1641 \\
            & $^3$F$_4$  & 45214 &  39564 & 41639 &  39563 & 1.1377 \\ 
            & $^3$H$_5$  & 59727 &  53798 &       &  53668 & 1.0333 \\
            & $^3$F$_2$  & 68538 &  61429 & 62930 &  62140 & 0.8331 \\
            & $^1$G$_4$  & 68885 &  62261 & 64588 &  62380 & 0.9902 \\ 
            & $^3$F$_3$  & 71917 &  65180 & 67154 &  65438 & 1.0833 \\ 
            & $^3$H$_4$  & 92224 &  84524 &       &  84662 & 0.9221 \\
            & $^1$D$_2$  & 98067 &  89273 & 90317 &  91341 & 1.1315 \\
            & $^1$J$_6$  &110065 & 101136 &       & 103487 & 1.0026 \\
\end{tabular}	
%\footnotetext[1]{Disputed interpretation, see text.}		
\end{ruledtabular}
\end{table}

Energy levels of the Ir$^{17+}$ ion are presented in Table~\ref{t:IrE}, M1 transition energies are in Table~\ref{t:Irwm1} while E1 transition energies are in Table~\ref{t:Irwe1}. In all these tables our results are compared with previous calculations and experiment. Note that experimental work~\cite{Ir17} presents only transition energies which are not translated into energy levels. This is because the interpretation of the E1 transition energies was ambiguous. 
Table~\ref{t:Irwe1} gives two alternative interpretations given in Ref.~\cite{Ir17}. Both calculations, present work and ~\cite{Ir17t} support first version.
Using this interpretation of the experimental data allows to reconstruct some experimental energy levels. The results are presented in Table~\ref{t:IrE}.
Note good agreement between latest calculations and between theory and experiment, while our old results~\cite{hci1} are less accurate.
There is one state  ($^3$F$_3$) for which the agreement between theory and experiment is poor. 
This leads to poor agreement for the transition energy involving this state (the $^1$D$_2$ - $^3$F$_3$ transition, see Table~\ref{t:Irwm1}). 
It is interesting to note that both theoretical results for this state agree well with each other but not with experiment.
It was noted in Ref.~\cite{Ir17t} that one possible reason for this might be wrong interpretation of experimental data.
However, we don't see any alternative interpretation. In the end, the reason for poor accuracy is not clear.
%Note however, that the accuracy for the energy tends to go down while moving up on the energy scale. This is because in present approach all states are divided into two groups, low-energy states and high-energy states. The contribution of the high energy states is included into the CI calculations perturbatively. As a result, the closer one gets to the border between these groups of the states the lower the accuracy.

%As it was noted in Ref.~\cite{Ir17t} one possible reason for this is wrong interpretation of experimental data. In Table~\ref{t:Irwm1} we present an alternative interpretation for this transition energy (the $^3$H$_5$ - $^3$H$_6$ transition) which is in much better agreement with calculations. Unfortunately, this transition is not connected to any other known experimental energy and thus does not lead to determination of any new energy levels. On the other hand, the experimental energy of the $^3$F$_3$ state remains unknown. 

Another  large deviation between theory and experiment is for the $^1$D$_2$ - $^3$F$_2$ transition (6.6\%).
Note however that the accuracy for each state is good, $\sim$ 1\% (see Table~\ref{t:IrE}). But it is -1\% for lower state and +1\% for upper state.
This leads to larger relative deviation in the difference. 

In the end the agreement between theory and experiment is within 3\% for all states and most of energy intervals. %up to 6\% for some states.
We expect similar accuracy for the Os$^{16+}$ ion.

\begin{table}
\caption{\label{t:Irwm1}
M1 transition energies in the Ir$^{17+}$ ion (in cm$^{-1}$). Comparison with previous calculations and experiment.
$\Delta E/E$ is the relative deviation of our results from the experimental energies (in \%).}
\begin{ruledtabular}
\begin{tabular}   {llcl rrr c}
\multicolumn{1}{c}{Config.}&
\multicolumn{3}{c}{Terms}&
\multicolumn{1}{c}{Expt.}&
\multicolumn{3}{c}{Theory}\\
&&&&\multicolumn{1}{c}{Ref.\cite{Ir17}}&
\multicolumn{1}{c}{Ref.\cite{Ir17t}}&
\multicolumn{1}{c}{Present}&
\multicolumn{1}{c}{$\Delta E/E$}\\
\hline
$4f^{13}5s$ & $^3$F$^{\rm o}_2$ &-& $^3$F$^{\rm o}_3$ & 20711 &  20409 & 20551 & -0.8 \\
            & $^1$F$^{\rm o}_3$ &-& $^3$F$^{\rm o}_4$ & 30359 &  30395 &   30168 & -0.6 \\
	                            
$4f^{12}5s^2$ & $^3$H$_5$&-&$^3$H$_6$  & 23640 &  23515 &   23973 & 1.4 \\
              & $^3$H$_4$&-&$^1$G$_4$  & 22430 &  22263 &   22282 & -0.7 \\
              & $^1$G$_4$&-&$^3$F$_4$  & 22949 &  22697 &   22817 & 1.7 \\
              & $^1$D$_2$&-&$^3$F$_3$  & 23163 &  24093 &   25903 & 12 \\
%              & $^3$H$_5$&-&$^3$H$_6$\footnotemark[2]  & 23163 &  23515 &   23973 & 3.5 \\
              & $^3$F$_3$&-&$^3$F$_4$  & 25515 &  25616 &   25875 & 1.4 \\
              & $^1$D$_2$&-&$^3$F$_2$  & 27387 &  27844 &   29201 & 6.6 \\
              & $^3$H$_4$&-&$^3$H$_5$  & 30798 &  30726 &   30994 & 0.6 \\
\end{tabular}
%\footnotetext[1]{Disputed interpretation.}		
%\footnotetext[2]{Alternative interpretation.}			
\end{ruledtabular}
\end{table}

\begin{table}
\caption{\label{t:Irwe1}
E1 transition energies in the Ir$^{17+}$ ion (in cm$^{-1}$). Comparison with the  previous calculations and experiment.}
\begin{ruledtabular}
\begin{tabular}   {lcl rrrr}
\multicolumn{3}{c}{Terms}&
\multicolumn{1}{c}{Ref.\cite{Ir17t}}&
\multicolumn{1}{c}{Present}&
\multicolumn{2}{c}{Expt.~\cite{Ir17}}\\
&&&&&\multicolumn{1}{c}{Version 1}&
\multicolumn{1}{c}{Version 2}\\
\hline
$^3$F$_4$&-&$^3$F$_4^{\rm o}$ & 39568 & 39563 & 41639 & \\
$^3$F$_3$&-&$^1$F$_3^{\rm o}$ & 34785 & 35271 & 36796 & 41639 \\
$^1$G$_4$&-&$^1$F$_3^{\rm o}$ & 31866 & 32213 &           & 39072 \\
\end{tabular}			
\end{ruledtabular}
\end{table}

\section{Calculations for the Os$^{16+}$ ion. Identification of the clock transitions.}

We have calculated energies, $g$-factors, lifetimes and other characteristics of few lowest states of Os$^{16+}$.
The results are presented in Table~\ref{t:OsE}. The list of states is very similar to those of the Ir$^{17+}$ ion (see previous section) but they go in different order.

Lifetimes ($\tau$) of even states are calculated by including all possible M1 and E2 transitions to low states. Lifetimes of odd states are calculated by taking into account E1 transitions to lower even states. We see at least two metastable states, the first excited state at E=9853~cm$^{-1}$ and $\tau$=1400~s, and the $^3$F$_2$ state at E=30675~cm$^{-1}$ and $\tau$=156~s. The first odd state at E=32908~cm$^{-1}$ is also relatively long-living, $\tau$=96~ms.
In principle, all these states can be used for high-precision measurements. Corresponding clock transitions within even configuration and $s-f$ transition are shown on Fig.~\ref{f:Os}.

 The E1 amplitude $\langle ^3{\rm F}^o_4 || E1 || ^3{\rm F}_4 \rangle = 1.91 \times 10^{-3}  a_B$ is small due to configuration mixing and compact ion size. %E.g., $\langle 5s|r|5s\rangle = 0.9  a_B$. 
The values of the E1 transition amplitudes for Ir$^{17+}$ are similar. For example, for three transitions presented in Table~\ref{t:Irwe1} the values are $2.3 \times 10^{-3}a_B$, $7.6 \times 10^{-4}a_B$ and $3.1 \times 10^{-3}a_B$ respectively.

To find the effect of black-body radiation (BBR) on clock transitions we have calculated static dipole polarizabilities of the low states of  Os$^{16+}$.
The calculation were performed by a method especially developed in our earlier work~\cite{symmetry} for atoms with open shells. 
The results are presented in Table~\ref{t:OsE} (static tensor polarizabilities $\alpha_2$ are also included).
The BBR shift (in Hz) is given by (see, e.g.~\cite{Derevianko})
\begin{equation}\label{e:BBR}
\delta \nu_{\rm BBR} = -8.611 \times 10^{-3}\left(\frac{T}{300K}\right)^4 \Delta \alpha_0.
\end{equation}
Using numbers from Table~\ref{t:OsE} one gets $\delta \nu/\nu \sim 10^{-20}$ for the E2 clock transitions and $\delta \nu/\nu \sim 10^{-17}$ for the $s-f$  transition.

All clock states of the Os$^{16+}$ ion have relatively large total angular momentum $J$. This means that the states might be sensitive to the gradient of electric field $\varepsilon$ via quadrupole interaction. Corresponding energy shift is given by
\begin{equation}\label{e:qu}
\Delta E_Q = \frac{J_z^2 - J(J+1)}{2J(2J-1)}Q\frac{\partial \varepsilon_z}{\partial z},
\end{equation}
where $Q$ is atomic quadrupole moment defined as doubled expectation value of the E2 operator in the stretched state
\begin{equation}\label{e:Q}
Q=2\langle J,J_z=J|E2|J,J_z=J\rangle.
\end{equation}
The calculated values of the quadrupole moment $Q$ for low states of Os$^{16+}$ are presented in Table~\ref{t:OsE}.

\begin{table}
\caption{\label{t:OsE}
Excitation energies ($E$), $g$-factors and other characteristics of low-lying states of the Os$^{16+}$ ion.
The numbers in second column are experimental energies reconstructed from transition energies presented in Ref.~\cite{Ir17} (supplemental material).
$\tau$ is the lifetime, $\alpha_0$ and $\alpha_2$ are the static scalar and tensor polarizabiities, $Q$ is the electric quadrupole moment.}
\begin{ruledtabular}
\begin{tabular}   {lrr ccc rr}
%\multicolumn{2}{c}{State}&
\multicolumn{1}{c}{Term}&
\multicolumn{2}{c}{$E$ (cm$^-1$)}&
\multicolumn{1}{c}{$g$}&
\multicolumn{1}{c}{$\tau$}&
\multicolumn{1}{c}{$\alpha_0$}&
\multicolumn{1}{c}{$\alpha_2$}&
\multicolumn{1}{c}{$Q$}\\
%\multicolumn{1}{c}{$F$}\\
&\multicolumn{1}{c}{\cite{Ir17}}&
\multicolumn{1}{c}{Present}&
&\multicolumn{1}{c}{$s$}&
\multicolumn{1}{c}{$a_0^3$}&
\multicolumn{1}{c}{$a_0^3$}&
\multicolumn{1}{c}{a.u.}\\
%\multicolumn{1}{c}{MHz/fm$^2$}\\
\hline
%                             E         g    tau          pol0      pol2      Q
\multicolumn{8}{c}{Even states, $4f^{12}5s^2$}\\
 $^3$H$_6$ &     0 &     0 & 1.164 &           & 0.5310 &-0.0085 &  0.194 \\
$^3$F$_4$ &  9049 &  9583 & 1.137 & 1.4[+4]   & 0.5307 & 0.0006 & -0.018 \\
$^3$H$_5$ & 21176 & 22433 & 1.033 & 4.0[-3]   & 0.5309 &-0.0074 &  0.171 \\
$^3$F$_2$ & 28951 & 30675 & 0.824 & 156       & 0.5304 & 0.0036 & -0.087 \\
$^1$G$_4$ & 29109 & 30643 & 0.989 & 10[-3]    & 0.5307 &-0.0033 &  0.075 \\
$^3$F$_3$ & 31931 & 33525 & 1.083 & 5.2[-3]   & 0.5305 & 0.0018 & -0.047 \\
$^3$H$_4$ & 49240 & 50086 & 0.927 &&&& \\
$^3$F$_2$ & 54221 & 58036 & 1.130 &&&& \\

%$^3$H$_6$ &     0 &  1.1643 &           & 0.5310 &-0.0085 &  0.194 \\
%$^3$F$_4$ &  9583 &  1.1370 & 1.4[+4]   & 0.5307 & 0.0006 & -0.018 \\
% $^3$H$_5$ & 22433 &  1.0333 & 4.0[-3]   & 0.5309 &-0.0074 &  0.171 \\
% $^1$G$_4$ & 30643 &  0.9894 & 10[-3]    & 0.5307 &-0.0033 &  0.075 \\
% $^3$F$_2$ & 30675 &  0.8239 & 156       & 0.5304 & 0.0036 & -0.087 \\
% $^3$F$_3$ & 33525 &  1.0833 & 5.2[-3]   & 0.5305 & 0.0018 & -0.047 \\
\multicolumn{8}{c}{Odd states, $4f^{13}5s$}\\
 $^3$F$^{\rm o}_4$ & & 32908 &  1.2500 & 96[-3]      & 0.0450 & 1.2[-5] &  0.191 \\
 $^3$F$^{\rm o}_3$ & & 37432 &  1.0524 & 38[-3]      & 0.0450 & 8.4[-5] &  0.161 \\
\end{tabular}			
\end{ruledtabular}
\end{table}

%Even:
%  1) En= 215858.36 J= 2.0 ME= -0.347331E+06  MHz  -439
%  1) En= 217668.29 J= 3.0 ME= -0.348341E+06  MHz -1449
%  1) En=  10235.15 J= 4.0 ME= -0.346802E+06  MHz    90
%  1) En=  22205.67 J= 5.0 ME= -0.348344E+06  MHz -1452  11619?
%  1) En=    858.17 J= 6.0 ME= -0.346892E+06  MHz   0  (GS)
%Odd:
%  1) En=  40636.29 J= 3.0 ME=  0.162368E+06  MHz 509260
%  1) En=  36110.85 J= 4.0 ME=  0.162434E+06  MHz 509326

\section{Search for new physics}

\subsection{Time variation of the fine structure constant}

High sensitivity to the variation of the fine structure constant $\alpha$ ($\alpha=e^2/\hbar c$) was the primary reason for suggesting the ions like Os$^{16+}$, Ir$^{17+}$ and others for the measurements~\cite{HCI,hci1}. Largest sensitivity correspond to largest change of the total electron angular momentum $j$ in the single-electron approximation for atomic transition~\cite{HCI}. Both ions, Os$^{16+}$ and Ir$^{17+}$, have optical transition between states of the $4f^{12}5s^2$ and $4f^{13}5s$ configurations, which correspond to the $s-f$ single-electron transition ($\Delta j$ = 2 or 3).

To find the sensitivity of atomic transitions to the variation of $\alpha$ we write the frequencies of the transitions in the form
\begin{equation}\label{e:q}
\omega_a(x) = \omega_{a0} + q_ax,
\end{equation}
where $x=\left( (\alpha/\alpha_0)^2 - 1 \right)$, $\alpha_0$ is physical value of $\alpha$ and $q$ is sensitivity coefficient to be found from calculations.
To find $q$  we vary the value of $\alpha$ in computer codes and calculate numerical derivative
\begin{equation}\label{e:qn}
q_a = \frac{\omega_a(\delta)-\omega_a(-\delta)}{2\delta}.
\end{equation}
Usually, we take $\delta=0.01$. Varying $\delta$ is useful for checking the stability of the results.

To search the manifestation of the variation of the fine structure constant we need at least two atomic transitions and measure one frequency against the other over a long period of time. Then the relative change of frequencies can be written as
\begin{equation}\label{e:K}
\frac{\delta \omega_a}{\omega_a} - \frac{\delta \omega_b}{\omega_b} = \left(\frac{2q_a}{\omega_a} - \frac{2q_b}{\omega_b}\right)\frac{\delta \alpha}{\alpha} \equiv \left(K_a - K_b\right)\frac{\delta \alpha}{\alpha},
\end{equation}
where dimensionless parameter $K$ ($K=2q/\omega$) is called enhancement factor. It is obvious from (\ref{e:K}) that for the highest sensitivity one needs two transitions with very different values of $K$, e.g., one is large and another is small or have opposite sign. The calculated values of $q$ and $K$ are presented in Table~\ref{t:alfa}.
The results for Os$^{16+}$ have been obtained in present work while the results for Ir$^{17+}$ are taken from~\cite{hci1}. 
One can see that $\Delta K \sim 20$ for transitions between different configurations, and $\Delta K \sim 1$ for transitions within one configuration.
For example, for two clock transitions of Os$^{16+}$ ($^3$H$_6$ - $^3$F$_4$ and $^3$F$_4$ - $^3$F$_4^{\rm o}$)  
\begin{equation}\label{e:KOs}
\frac{\delta \omega_a}{\omega_a} - \frac{\delta \omega_b}{\omega_b} \approx 25 \frac{\delta \alpha}{\alpha}.
\end{equation}
Note that it looks beneficial to search for transitions with small $\omega$ for the sake of having large enhancement factor ($K=2q/\omega$).
However, sometimes such transitions  have no advantage since accuracy of the measurements is equally important and the ratio of the relative experimental uncertainty to the relative change of $\omega$ due to variation of $\alpha$ often does not depend on $\omega$  ($(\delta \omega_{expt}/\omega)/(\delta \omega_{\alpha}/\omega) = \delta \omega_{expt}/\delta \omega_{\alpha}$) - see Ref.~\cite{cipt1} for a detailed discussion). Large values of $q$ in HCI give real advantage. 

\begin{table}
\caption{\label{t:alfa}
Parameters of the Os$^{16+}$ and Ir$^{17+}$ ions relevant to the search for the variation of $\alpha$.
The values of the $q$-coefficients for the Ir$^{17+}$ ion are taken from Ref.~\cite{hci1}.
The enhancement factor $K$ is given by $K=2q/E$.}
\begin{ruledtabular}
\begin{tabular}   {ll rr c}
\multicolumn{2}{c}{State}&
%\multicolumn{1}{c}{Term}&
\multicolumn{1}{c}{$E$}&
\multicolumn{1}{c}{$q$}&
\multicolumn{1}{c}{$K$}\\
&&\multicolumn{1}{c}{[cm$^{-1}$]}&
\multicolumn{1}{c}{[cm$^{-1}$]}&\\
\hline
%New physics                  E       q       K     P2Pz   D1 D2                 
\multicolumn{5}{c}{Os$^{16+}$, even states}\\
$4f^{12}5s^2$ & $^3$H$_6$ &     0 &     0 &     0  \\
              & $^3$F$_4$ &  9583 & -1427 & -0.30  \\
	      & $^3$H$_5$ & 22433 & 19401 &  1.73  \\
	      & $^1$G$_4$ & 30643 & 20900 &  1.36  \\
	      & $^3$F$_2$ & 30675 &  6870 &  0.45  \\
	      & $^3$F$_3$ & 33525 & 19300 &  1.15  \\
\multicolumn{5}{c}{Os$^{16+}$, odd states}\\
$4f^{13}5s$   & $^3$F$^{\rm o}_4$ & 32908 & 337726 & 23.9  \\
              & $^3$F$^{\rm o}_3$ & 37432 & 339746 & 20.8  \\ 
\multicolumn{5}{c}{Ir$^{17+}$, odd states}\\
$4f^{13}5s$ & $^3$F$^{\rm o}_4$ &      0 &     0 &  0    \\
            & $^3$F$^{\rm o}_3$ &   4647 &  2065 & 0.9 \\
            & $^3$F$^{\rm o}_2$ &  25198 & 24183 & 1.9 \\
            & $^1$F$^{\rm o}_3$ &  30167 & 25052 & 1.7  \\
\multicolumn{5}{c}{Ir$^{17+}$, even states}\\
%Even states   	                            
$4f^{12}5s^2$& $^3$H$_6$  & 29695 & -385367 & -26   \\
             & $^3$F$_4$  & 39563 & -387086 & -20   \\
             & $^3$H$_5$  & 53668 & -362127 & -13   \\
             & $^3$F$_2$  & 62140 & -378554 & -12  \\
             & $^1$G$_4$  & 62380 & -360678 & -12  \\
             & $^3$F$_3$  & 65438 & -362313 & -11  \\
             & $^3$H$_4$  & 84662 & -339253 &  -8  \\
             & $^1$D$_2$  & 91341 & -363983 &  -8  \\
             & $^1$J$_6$ & 103487 & -364732 &  -7  \\
\end{tabular}			
\end{ruledtabular}
\end{table}

\subsection{Einstein equivalence principle violation}

Local position invariance (LPI), local Lorentz invariance (LLI), and the weak equivalence principle form the Einstein equivalence principle, which is the foundation of general relativity. Some extensions of the Standard Model allow for violation of these invariances.
The LPI violating term can be written as (see e.g. \cite{LPI1} and references therein)
\begin{equation}\label{e:LPI}
\hat H_{\rm LPI} = C_{00}\frac{2}{3}\frac{U}{c^2}\hat K,
\end{equation}
where $C_{00}$ is an unknown constant , $U$ is the gravitation potential, $c$ is the speed of light and $\hat K$ is the operator of kinetic energy which in relativistic case can be written as $\hat K = c\gamma_0\gamma^i p_i/2$, $\mathbf{p} = -i\hbar \mathbf{\nabla}$ is the operator of electron momentum.

The presence of term (\ref{e:LPI}) in the Hamiltonian causes the change of the atomic frequencies due to  the change of gravitation potential $U$ (e.g., due to the annual variation of the Sun-Earth distance). It can be shown using virial theorem that in the non-relativistic limit all atomic frequencies change at the same rate and the effect is not detectable~\cite{RRR}.  Therefore, it is convenient to describe the effect in terms of the so called relativistic factor $R$ which indicates the deviation from the non-relativistic virial theorem in the relativistic case~\cite{RRR}
\begin{equation}\label{e:R}
R_{ab} = -\frac{E_{Ka}-E_{Kb}}{E_a-E_b},
\end{equation}
where $E_{Ka}$ is the kinetic part of the energy of atomic state $a$ and $E_a$ is its full energy.
Then the relative change of two atomic frequencies can be written as
\begin{equation}\label{e:RR}
\frac{\Delta \omega_{ab}}{\omega_{ab}} - \frac{\Delta \omega_{cd}}{\omega_{cd}} = - (R_{ab}-R_{cd})\frac{2}{3}c_{00}\frac{\Delta U}{c^2}.
\end{equation}
The highest sensitivity of atomic frequencies to the variation of gravitation potential $U$ can be achieved for transitions between states with very different values of the relativistic factor $R$. It turns out that similar to the case of variation of $\alpha$ the highest sensitivity is for the transitions between states of different configurations.

To calculate the values of $R$ one needs to calculate kinetic energies $E_K$ caused by the  kinetic energy operator $\hat K$. In principle, one can use the standard approach based on the RPA equations (\ref{e:RPA}) and calculating matrix elements (\ref{e:Tij}). However, calculations of the matrix elements of the kinetic energy operator are very sensitive to the correlation effects  and one needs to include many minor contributions for stable results.
It is more practical to use the so called finite field approach in which the calculation of the energy shift caused by a scalar operator is reduced to the calculation of the energy. The operator is added to the Dirac equations with a rescaling operator $s$, calculations are repeated several times for different but small values of $s$ and then extrapolated to $s=1$.
In our case, the Dirac equations with rescaled operator of kinetic energy can be written as
\begin{eqnarray}\
&& \left(\frac{\partial f}{\partial r} + \frac{\kappa}{r}f\right)(1+s) - \left[1+\alpha^2(\epsilon -\hat V)\right]g=0 \nonumber \\
&& \left(\frac{\partial g}{\partial r} - \frac{\kappa}{r}g\right)(1+s) + (\epsilon -\hat V)f=0 \label{e:Ds}.
\end{eqnarray}
We perform calculation for several values of $s$ from $s=0$ to $s=10^{-5}$, extrapolate the results to $s=1$ to get kinetic energies $E_K$, and use (\ref{e:R}) to get the values of $R$. The results for the Os$^{16+}$ and Ir$^{17+}$ ions are presented in Table~\ref{t:NP}. Calculations show that the energy shift caused by the kinetic energy operator $\hat K$ is similar for all states of the same configuration. Different values of $R$ are mostly due to different energy intervals in the denominator (\ref{e:R}). The values of $\Delta R$ are presented with respect to the ground state. Therefore, the values are relatively small for all states of the ground state configuration.
On the other hand,  the values of $\Delta R$ are large for the transitions between states of different configurations. Here $R \gg 1$, which is probably common for all HCI. In contrast, $R \sim 1$ for neutral atoms~\cite{cipt3}. Note also that $\Delta R$ for transitions between states of different configurations of Os$^{16+}$ and Ir$^{17+}$ ions have different signs. This is because of the different order of states in two ions.

The LLI violating term can be written 
\begin{equation}\label{e:LLI}
\hat H_{\rm LLI} = -\frac{1}{6}C_0^{(2)}T_0^{(2)},
\end{equation}
where $T_0^{(2)}$ is a tensor operator $T_0^{(2)} =c\gamma_0(\gamma^jp_j-3\gamma^3p_3)$.  
The presence of term (\ref{e:LLI}) in the Hamiltonian leads to the dependence of atomic frequencies on the orientation of the apparatus in space.
For the interpretation of the measurements one needs to know the values of the reduced matrix elements of the operator $T_0^{(2)}$.
We perform the calculations for the Os$^{16+}$ and Ir$^{17+}$ ions using standard approach described in section \ref{s:RPA}.
The results are presented in Table~\ref{t:NP}.

It was stated in Ref.~\cite{Shaniv,Nature} that to study the LLI violation one could measure the frequency of the transitions between states with different values of the projection of the total electron momentum $J_z$ within one metastable state. Large value of the matrix element of the $T_0^{(2)}$ operator and long lifetime of the state are needed for high sensitivity. In Os$^{16+}$ and Ir$^{17+}$ we have large values of the matrix element in many states including ground state. 
In both ions the values are larger than in the Yb$^{+}$ ion suggested for the most sensitive measurements in Ref.~\cite{Nature}. These makes the ions to be attractive candidates for the study of the LLI violation.

%\newpage

\begin{table}
\caption{\label{t:NP}
Parameters of the Os$^{16+}$ and Ir$^{17+}$ ions relevant to the search for the Einstein equivalence principle violation.
The values of the relativistic factors $R$ are presented with respect to the ground state ($\Delta R_{ag} = R_a -R_g$).}
\begin{ruledtabular}
\begin{tabular}   {ll rr c }
\multicolumn{2}{c}{State}&
%\multicolumn{1}{c}{Term}&
\multicolumn{1}{c}{$E$}&
\multicolumn{1}{c}{$\Delta R$}&
\multicolumn{1}{c}{$\langle v|| T^{(2)}||v\rangle$}\\
%\multicolumn{1}{c}{$D_1$}&
%\multicolumn{1}{c}{$D_2$}\\
&&\multicolumn{1}{c}{[cm$^{-1}$]}&&
\multicolumn{1}{c}{[a.u.]} \\
%\multicolumn{2}{c}{[GHz]} \\
\hline
%New physics                  E    R  P2Pz                   
\multicolumn{5}{c}{Os$^{16+}$, even states}\\
$4f^{12}5s^2$ & $^3$H$_6$ &     0 & 0 & -299  \\
              & $^3$F$_4$ &  9583 & 0.2 &   24  \\
	      & $^3$H$_5$ & 22433 & 7 & -256  \\
	      & $^1$G$_4$ & 30643 & 5 & -106  \\
	      & $^3$F$_2$ & 30675 & 2 &  122  \\
	      & $^3$F$_3$ & 33525 & 5 &   68  \\
			            
\multicolumn{5}{c}{Os$^{16+}$, odd states}\\
$4f^{13}5s$   & $^3$F$^{\rm o}_4$ & 32908 & 18 & -272   \\
              & $^3$F$^{\rm o}_3$ & 37432 & 17 & -223  \\ 
\multicolumn{5}{c}{Ir$^{17+}$, odd states}\\
$4f^{13}5s$ & $^3$F$^{\rm o}_4$ &      0 & 0 & -283   \\
            & $^3$F$^{\rm o}_3$ &   4647 & 3 & -233  \\
            & $^3$F$^{\rm o}_2$ &  25198 & 3 & -197  \\
            & $^1$F$^{\rm o}_3$ &  30167 & 3 & -254  \\
\multicolumn{5}{c}{Ir$^{17+}$, even states}\\
%Even states   	                            
$4f^{12}5s^2$& $^3$H$_6$  & 29695 & -30 & -311  \\
             & $^3$F$_4$  & 39563 & -20 &   26  \\
             & $^3$H$_5$  & 53668 & -13 & -266   \\
             & $^3$F$_2$  & 62140 & -11 &  131  \\
             & $^1$G$_4$  & 62380 & -11 & -107 \\
             & $^3$F$_3$  & 65438 & -10 &   71  \\
             & $^3$H$_4$  & 84662 & -7 & -138  \\
             & $^1$D$_2$  & 91341 & -7 &   95  \\
             & $^1$J$_6$ & 103487 & -5 & -612  \\
\end{tabular}			
\end{ruledtabular}
\end{table}

\subsection{Search for new bosons using non-linearities of King plot.}

\begin{table}
\caption{\label{t:King}
Field shift constants $F$ and energy shifts $D$ due to the Yukawa-type electron-nucleon interaction.
$D_1$ is calculated at $m_{\phi}$=3~MeV, $D_2$ is calculated at $m_{\phi}$=0.3~MeV.
All values are given with respect to the ground state.}
\begin{ruledtabular}
\begin{tabular}   {ll rr c rr}
\multicolumn{2}{c}{State}&
%\multicolumn{1}{c}{Term}&
\multicolumn{1}{c}{$E$}&
\multicolumn{1}{c}{$F$}&
%\multicolumn{1}{c}{$\langle v|| T^{(2)}||v\rangle$}&
\multicolumn{1}{c}{$D_1$}&
\multicolumn{1}{c}{$D_2$}\\
&&\multicolumn{1}{c}{[cm$^{-1}$]}&
\multicolumn{1}{c}{[MHz/fm$^2$]} &
\multicolumn{2}{c}{[GHz]} \\
\hline
%New physics                  E    F   D1   D2          
%New physics                  E    F   D1   D2          
\multicolumn{6}{c}{Even states}\\
$4f^{12}5s^2$ & $^3$H$_6$ &     0 &     0  & 0 & 0 \\
              & $^3$F$_4$ &  9583 &    93  &  -0.010 & -0.177 \\
	      & $^3$H$_5$ & 22433 & -1487  &  0.157 & 2.86 \\
	      & $^1$G$_4$ & 30643 & -1612  &  & \\
	      & $^3$F$_2$ & 30675 &  -564  &  0.047 & 0.879 \\
	      & $^3$F$_3$ & 33525 & -1487  &  0.157 & 2.86 \\
			            
\multicolumn{6}{c}{Odd states}\\
$4f^{13}5s$   & $^3$F$^{\rm o}_4$ & 32908 & 5.32[+5] & -52.3 & -490 \\
              & $^3$F$^{\rm o}_3$ & 37432 & 5.32[+5] & -52.3 & -490 \\ 
\end{tabular}			
\end{ruledtabular}
\end{table}

It was suggested in Ref.~\cite{KP1,KP2} that non-linearities of King plot can be used to put limits on new interactions.
Isotope shift $\nu$ for a specific atomic transition $a$ can be written in a simliest form as
\begin{equation}\label{e:IS}
\nu_a = F_a \delta \langle r^2 \rangle_{ij} + K_a \mu_{ij}.
\end{equation}
Here $F_a$ is the field shift constant, $K_a$ is the mass shift constant, $\delta \langle r^2 \rangle_{ij}$ is the change of the root-mean-square nuclear radius between isotopes $i$ and $j$, $\mu_{ij} =1/m_i-1/m_j$ is the reduced mass of the two isotopes. Here we neglect higher-order in nuclear structure terms (e.g., terms $\sim  \delta \langle r^2 \rangle_{ij}^2$, $\delta \langle r^4 \rangle_{ij}$, etc.). If we have two transitions, then finding $\delta \langle r^2 \rangle_{ij}$ from (\ref{e:IS}) and substituting it to similar equation for another transition leads to
\begin{equation}\label{e:King}
\tilde \nu_{aij} = \frac{F_a}{F_b} \tilde \nu_{bij} - \frac{F_a}{F_b} K_b+K_a,
\end{equation}
where $\tilde \nu = \nu/\mu$. This equation presents a straight line on the ($\tilde \nu_a$, $\tilde \nu_b$) plane (King plot).
At least two transitions and four isotopes are needed to see  deviations from the straight line. 
These conditions are satisfied for the Os$^{16+}$ ion. 
Extra terms in (\ref{e:IS}) can lead to non-linearities of King plot.
For example, if we have extra electron-neutron interaction mediated by a scalar boson of mass $m_{\phi}$ via Yukawa-type interaction, then there is a contribution to the isotope shift due to different number of neutrons in two isotopes. Corresponding extra term can be written as $\frac{\alpha_{\rm NP}}{\alpha}\Delta N_{ij} D_a$, where $\alpha_{\rm NP}$ is a dimensionless constant of the strength of new interaction, $\alpha$ is the fine structure constant, $\Delta N_{ij}$ is the difference in the number of neutrons in isotopes $i$ and $j$, and
$D=\langle \exp{(-m_{\phi}cr/\hbar)} \rangle$.  Then Eq.~(\ref{e:King}) becomes
\begin{equation}\label{e:King-NP}
\tilde \nu_{aij} = \frac{F_a}{F_b} \tilde \nu_{bij} - \frac{F_a}{F_b} K_b+K_a + D_b\gamma_{ij}\left(\frac{F_a}{F_b}-\frac{D_a}{D_b}\right).
\end{equation}
Here $\gamma_{ij} = \frac{\alpha_{\rm NP}}{\alpha} \Delta N/\mu_{ij}$.
The last term in (\ref{e:King-NP}) depends on the isotopes and therefore may break the linearity of King plot. Studying possible non-linearities puts limit on the value of $\alpha_{\rm NP}$ if the values of $F$ and $D$ are known. We calculate these values for different states of Os$^{16+}$ using technique described in section \ref{s:RPA}. The results are presented in Table~\ref{t:King}. The value of $D$ depends on the mass of extra boson $m_{\phi}$. For  mass $m_{\phi} > 30$~MeV,  radius of new interaction is smaller than  the nuclear radius  $R_N=\hbar/m_{\phi}c=6.5$~fm) and the new interaction is  indistinguishable from the field shift. In this case $D_a/D_b = F_a/F_b$ and the last term in Eq.  (\ref{e:King}) vanishes.
We calculate $D$ for two values of $m_{\phi}$,  3 MeV and 0.3 MeV,  which correspond to the new interaction radii $r=10R_N$ and $r=100R_N$. For these values of $D$ the last term in Eq. (\ref{e:King}) is not zero and can be used to put limits on $\alpha_{\rm NP}$ at given $m_{\phi}$.

Note that higher-order  nuclear structure terms can also lead to the non-linearities of the King plot. For example, it was demonstrated in Ref.~\cite{Yb+} that observed non-linearities of the King plot in Yb$^+$ can be explained by significant variation of the nuclear deformation between Yb isotopes (which produces terms $\sim \delta \langle r^4 \rangle$). Unfortunately,  
these higher-order terms cannot be calculated with the accuracy  exceeding the accuracy of the measurements of the King plot nonlinearity in Yb$^+$.  As a result, the limits on the new interaction  have  been  obtained  under assumption that this new interaction is the only source of nonlinearities. %The best outcome for the measurements can be obtained if highly accurate measurements produce non non-linearities. In this case the limit on new interaction would be strong. In contrast, if the non-linearities are observed, then the most likely source for them comes from within the standard model. 
Therefore,  we do not consider higher-order terms in this work since their calculations  are unreliable and they will be neglected anyway. 
However,  corresponding non-linearities are likely to be smaller  for Os than that for Yb. This is because nuclear calculations suggest that the nuclear deformation for all stable even isotopes of Os are about the same (see Table~\ref{t:OsIs} and Ref.~\cite{ddpc}). Equal values of the nuclear quadrupole deformation $\beta$ produce equal energy shifts and no non-linearities~\cite{Yb+}. 

\section{Conclusion}

We have studied in detail electronic structure of the Ir$^{17+}$ and Os$^{16+}$ ions using advanced theoretical techniques.
Good agreement with available experimental data and earlier most advanced calculations is achieved.
%Some minor amendments in the interpretation of the experimental data for Ir$^{17+}$ bring them to very good agreement with the calculations.
Calculations reveal many useful features of both ions relevant to their use for very accurate optical clocks sensitive to new physics. The  Os$^{16+}$ ion has at least three long-living states and three transitions which are good candidates for clock transitions.
One of the transition is very sensitive to the variation of the fine structure constant. Many states of the ion, including ground state, are sensitive to the local Lorentz invariance and local position invariance violation. The latter feature  is likely to be common for all HCI with open $4f$ or $5f$ shells. In addition, the Os$^{16+}$ ion can be used to study new electron-neutron interactions using King plot and its possible non-linearities.

\acknowledgments

The authors are grateful to Hendrik Bekker for careful reading of the manuscript and many useful comments.
This work was supported by the Australian Research Council Grants No. DP230101058 and DP200100150.

\end{document}